\documentclass[aps,prd,preprint,superscriptaddress,tightenlines,nofootinbib]{revtex4}

\usepackage{graphicx}% Include figure files
\usepackage{dcolumn}% Align table columns on decimal point
\usepackage{bm}% bold math

\begin{document}

\preprint{CLNS 04/1884}       % for CLNS notes
\preprint{CLEO 04-09}         % for CLNS notes

\title{Observation of  $1^-0^-$
Final States from {\boldmath$\psi(2S)$} Decays and $e^+e^-$ Annihilation}

\author{N.~E.~Adam}
\author{J.~P.~Alexander}
\author{K.~Berkelman}
\author{D.~G.~Cassel}
\author{J.~E.~Duboscq}
\author{K.~M.~Ecklund}
\author{R.~Ehrlich}
\author{L.~Fields}
\author{R.~S.~Galik}
\author{L.~Gibbons}
\author{B.~Gittelman}
\author{R.~Gray}
\author{S.~W.~Gray}
\author{D.~L.~Hartill}
\author{B.~K.~Heltsley}
\author{D.~Hertz}
\author{L.~Hsu}
\author{C.~D.~Jones}
\author{J.~Kandaswamy}
\author{D.~L.~Kreinick}
\author{V.~E.~Kuznetsov}
\author{H.~Mahlke-Kr\"uger}
\author{T.~O.~Meyer}
\author{P.~U.~E.~Onyisi}
\author{J.~R.~Patterson}
\author{D.~Peterson}
\author{J.~Pivarski}
\author{D.~Riley}
\author{J.~L.~Rosner}
\altaffiliation{On leave of absence from University of Chicago.}
\author{A.~Ryd}
\author{A.~J.~Sadoff}
\author{H.~Schwarthoff}
\author{M.~R.~Shepherd}
\author{W.~M.~Sun}
\author{J.~G.~Thayer}
\author{D.~Urner}
\author{T.~Wilksen}
\author{M.~Weinberger}
\affiliation{Cornell University, Ithaca, New York 14853}
% ...
%\author{(CLEO Collaboration)} %FOR PRD_SPECIAL_CHANGEME
\collaboration{CLEO Collaboration} %FOR PRL,CLNS
\noaffiliation

\date{July 14, 2004}

\begin{abstract} 
Using CLEO data collected from CESR $e^+e^-$ collisions at the
$\psi(2S)$ resonance and nearby continuum at $\sqrt{s}$=3.67~GeV, 
we report the first significantly non-zero measurements of light
vector-pseudoscalar hadron pair production (including $\rho\pi$, $\omega\pi$,
$\rho\eta$, and $K^{*0}\bar{K^0}$) and the
$\pi^+\pi^-\pi^0$ final state, 
both from $\psi(2S)$ decays and direct $e^+e^-$ annihilation.
\end{abstract}

\maketitle

  The $\rho\pi$ puzzle poses one of the most enduring questions
in strong interaction physics: why is the branching fraction
for $\psi(2S)\to\rho\pi$ at least twenty~\cite{PDG} times smaller
than expected from scaling the $J/\psi\to\rho\pi$ rate by the ratio
of dilepton branching fractions? The ``12\% rule'', a scaling
conjecture generalizing this question for any 
decay mode,
has as its underlying assumption that since charmonium decay 
to light hadrons must proceed
through annihilation of the constituent $c\bar{c}$ into a photon or three
gluons, the decay width should be proportional to the square of the 
$c\bar{c}$ wave function overlap at the origin. 
The rule's figure of merit is
\begin{equation}
Q_X = {{{\cal B}(\psi(2S)\to X)/{\cal B}(J/\psi\to X)}
\over{{\cal B}(\psi(2S)\to \ell^+\ell^-)/{\cal B}(J/\psi\to
\ell^+\ell^-)}} \ \ ,
\end{equation}
where ${\cal B}$ denotes a branching fraction and $X$ a particular final
state. Decays to dileptons also proceed via $c\bar{c}$ annihilation, and
their branching fractions are well-measured~\cite{PDG}, 
so their ratio makes a suitable denominator 
in Eq.~(1). Several channels have $Q_X\approx 1$~\cite{PDG}, although some deviations 
from unity are expected~\cite{GULI}. The $\rho\pi$ mode is not alone: significant
suppressions also exist for at least one other vector-pseudoscalar 
(VP) channel ($K^*(892)^+K^-$) and three vector-tensor 
channels ($\rho a_2(1320)$, $K^*(892) \bar{K}_2^*(1430)$, and 
$\omega f_2(1270)$)~\cite{PDG,BESVT}. 

 The continuing struggle to understand the pattern of $Q$-values
has provoked many theoretical explanations. 
 For isospin-violating (IV) modes such as $\omega\pi$ and 
$\rho\eta$, three-gluon mediated decay is
suppressed, allowing 
the electromagnetic process of annihilation into a virtual photon to dominate.
Whether $Q_{\rm IV}\sim 1$ remains
a crucial open question.
A recent review~\cite{GULI} of relevant theory and experiment 
concludes that none of the proffered
theoretical explanations is satisfactory and also finds the
underpinnings of the 12\% rule overly simplistic.

  A major impediment to addressing the puzzle in a systematic manner 
is the dearth of
$\psi(2S)$ branching fraction measurements. 
Experimental progress on key VP final states has remained dormant 
for many years. Continuum production, 
$e^+e^-\to\gamma^*\to X$, which is of interest in its own 
right~\cite{CONTIN,BRODLEP},
is expected at levels that may affect $\psi(2S)$ backgrounds
and will interfere~\cite{WMYINTER} with $\psi(2S)$ decay,
but has not yet been
measured. Using $e^+e^-$ collision data acquired with
the CLEO detector operating at the Cornell Electron Storage Ring (CESR),
this Letter presents 
$\psi(2S)$ branching fractions and continuum cross sections for
$\pi^+\pi^-\pi^0$;
$\rho\pi$, $\omega\pi$, $\phi\pi$, $\rho\eta$, $\omega\eta$, $\phi\eta$,
$K^{*0}(892)\bar{K^0}$, $K^{*+}(892)K^-$; $b_1(1235)\pi$.
Where applicable, the inclusion of charge conjugate states is implied.
We use $\rho\to\pi\pi$, $\pi^0\to\gamma\gamma$, 
$\omega\to\pi^+\pi^-\pi^0$, $\phi\to K^+K^-$,
$\eta\to\gamma\gamma$ and $\pi^+\pi^-\pi^0$,
 $K^{*0}\to K^-\pi^+$,
$K^{*+}\to K_S^0\pi^+$ and $K^+\pi^0$, and $K_S^0\to\pi^+\pi^-$.

The CLEO~III detector~\cite{cleoiiidetector} 
features a solid angle
coverage of 93\% for charged and neutral particles.
For the data presented here, the charged particle tracking system operates in a
1.0~T magnetic field along the beam axis and achieves 
a momentum resolution of $\sim$0.6\% at
$p=1$~GeV/$c$. The cesium iodide (CsI) calorimeter attains photon 
energy resolutions of 2.2\% at $E_\gamma=1$~GeV and 5\% at 100~MeV.
Two particle identification systems, one based on ionization energy loss ($dE/dx$) in 
the drift chamber and the other a ring-imaging Cherenkov (RICH) 
detector, are used together to separate $K^\pm$ from $\pi^\pm$. 
The combined $dE/dx$-RICH particle identification has
efficiencies $>$90\% and misdentification rates $<$5\%
for both $\pi^\pm$ and $K^\pm$.

Half of the $\psi(2S)$ data and all
the $\sqrt{s}$=3.67~GeV data were taken after
a transition to CLEO-c~\cite{YELLOWBOOK}, in which
CLEO~III's silicon-strip vertex detector was replaced with a six-layer
all-stereo drift chamber. 
The two detector configurations correspond
to different accelerator lattices: the former with
a single wiggler magnet and a center-of-mass
energy spread $\Delta E$=1.5~MeV, the latter 
(CESR-c~\cite{YELLOWBOOK}) with the first
half of its full complement (12) of wiggler magnets and $\Delta E$=2.3~MeV. 

 The integrated luminosity ($\cal{L}$) of the datasets was measured 
using  $e^+e^-\to\gamma\gamma$
events~\cite{LUMINS}. Event counts were normalized with 
a Monte Carlo (MC) simulation based on the Babayaga~\cite{BBY} event 
generator combined with GEANT-based~\cite{GEANT} detector modeling. 
The datasets have $\cal{L}$=5.63~pb$^{-1}$ on the 
peak of the $\psi(2S)$ (2.74~pb$^{-1}$ for CLEO~III, 
2.89~pb$^{-1}$ for CLEO-c) and 20.46~pb$^{-1}$ at $\sqrt{s}$=3.67~GeV
(all CLEO-c). The scale factor applicable to continuum yields
in order to normalize them to $\psi(2S)$ data, $f=0.268\pm0.004$, includes a
2.6\% correction to the $\cal{L}$ ratio
to scale it by 1/$s^3$~\cite{BRODLEP}; the error includes both the 
relative luminosity and form factor $s$-dependence uncertainties.
We also correct each final state's $f$ for small 
efficiency differences between the $\psi(2S)$ and continuum
samples caused by detector configuration.

We base our event selection on charged particles 
reconstructed in the tracking system and photon candidates in the 
CsI calorimeter. 
Energy and momentum conservation is
required of the reconstructed hadrons, which have momenta $p_i$ and 
total energy $E_{\rm vis}$. 
We demand $0.98 < E_{\rm vis}/\sqrt{s} < 1.015$
and $| |p_1|-|p_2| |/(\sqrt{s}/c) < 0.02$
(for $\pi^+\pi^-\pi^0$, $p_1=p_{\pi^0}$ and $p_2=p_{\pi^+} + p_{\pi^-}$),
 which together suppress
backgrounds with missing energy 
or incorrect mass assignments. 
The experimental resolutions are
smaller than 1\% in scaled energy and 2\% in scaled momentum
difference. In order to suppress hadronic transitions to $J/\psi$, we reject
events in which any of the following fall within 3.05-3.15~GeV: the invariant mass of
the two highest momentum tracks; or the recoil mass from the lowest
momentum single $\pi^0$, $\pi^0\pi^0$ pair, or $\pi^+\pi^-$ pair.
Feeddown from $\pi^0\pi^0 J/\psi$, $J/\psi\to\mu^+\mu^-$
into $\pi^+\pi^-\pi^0$, $\rho^+\pi^-$, or $(K^+\pi^0)K^-$ is additionally 
suppressed by requiring $M(\mu^+\mu^-)<3.05$~GeV for those channels.

 MC studies were used to determine invariant mass windows for
intermediate particle decay products.
To reduce contamination from $\omega f_2(1270)$~\cite{BESVT}
and $\omega f_0(600)$~\cite{BESOMEGAPIPI} in $b_1\pi$, we exclude
$M_{\pi\pi}<$1.5~GeV. Similarly, $\rho\eta$ candidates with low mass $\eta\pi^\pm$ states
are avoided with $M(\eta\pi^\pm)_{\rm min}>1.4$~GeV. 
For $\pi^0\to\gamma\gamma$,
$\eta\to\gamma\gamma$, and $K_S^0\to\pi^+\pi^-$ candidates
we use kinematically constrained fits of the decay products to the 
parent masses. 
Fake $\pi^0$ and $\eta$ mesons are suppressed with lateral shower profile
restrictions and by requiring that
their decays to $\gamma\gamma$ not be too asymmetric. 

For $\pi^+\pi^-\pi^0$, $\rho^+\pi^-$, and $\rho\eta$ ($\phi\eta$) with $\eta\to\gamma\gamma$,
one of the two final state charged particles must be positively identified
as a $\pi^\pm$ ($K^\pm$), but neither can be positively identified as a $K^\pm$
($\pi^\pm$). Charged kaons in $K^*K$ must be 
identified as such, and any $\pi^\pm$ candidate must not be identified
as $K^\pm$. Charged particles 
must not be identified as electrons using criteria based on 
momentum, calorimeter energy deposition, and
$dE/dx$.
The softer charged particle in two-track modes 
must have $p<0.425\times \sqrt{s}/c$ to suppress potential background
from $\mu^+\mu^-\gamma$ in which a fake $\pi^0$ is found. 
Both tracks in two-track modes must satisfy
$|\cos\theta|<0.83$, where $\theta$ is the polar 
angle with respect to the $e^+$ direction.

The efficiency $\epsilon$ for each final state is the average obtained from MC
simulations~\cite{GEANT} of both detector configurations.
The VP modes are generated~\cite{EVTGEN,PHOTOS} with 
angular distribution $(1+\cos^2\theta)$~\cite{BRODLEP}, $b_1\pi$ flat
in $\cos\theta$, and $\pi^+\pi^-\pi^0$ 
as in $\omega$ decay. We assume ${\cal B}(b_1\to\omega\pi)$=100\%.

  Background contamination from other $\psi(2S)$ decays is determined
from sidebands neighboring the signal windows in $\pi^0$,
$\eta$, $\omega$, $\phi$, $K_S^0$, $K^*$, and $b_1$ 
candidate mass distributions.
The sideband yields from the  $\psi(2S)$ sample
are decremented by the corresponding number of scaled continuum events (because
scaled continuum events inside the signal window are subtracted separately) 
and by the small residual signal contributions expected, and then
scaled to match the signal window size.

  We normalize the branching fractions to the
total number of produced $\psi(2S)$ events.
The technique described in Ref.~\cite{ATHAR} is applied to
the datasets used here, resulting in a total number of
$\psi(2S)$ decays of 3.08$\times 10^6$.

 Kinematic distributions are shown
in Figs.~1-4 and the event totals and efficiencies
in Table~I. We observe signals for several modes
in both $\psi(2S)$ and continuum datasets. The
significances $S$ in the last column of Table~I
reflect the likelihood that the $\psi(2S)$ yields
cannot be attributed to backgrounds alone.
$S$ is computed from trials in which Poisson fluctuations of the 
$\psi(2S)$, continuum, and cross-feed contributions are all
simulated to obtain a confidence level (CL) that a given mean $\psi(2S)$
signal $\mu$ combined with 
backgrounds would exceed or equal the observed event count.
$S$ is obtained from this procedure with $\mu$=0.

Table~II shows the final results. We compute branching fractions with
a straightforward subtraction of luminosity-scaled continuum yields;
the value of the true branching fraction depends on the unknown
three-gluon decay amplitudes and corresponding unknown phases.
Statistical errors shown correspond to 68\% CL
and upper limits to 90\% CL, and are obtained through
simulated trials as described above.
Values of $Q$ are computed for each mode based on
branching fractions from Ref.~\cite{PDG}, except for 
${\cal B}(J/\psi\to\pi^+\pi^-\pi^0)=(2.10\pm0.12)\%$~\cite{BESRHOPI}.
Born-level cross sections at $\sqrt{s}$=3.67~GeV are also given
and include 
an upward adjustment of 20\% to account for 
initial state radiation~\cite{BBY}.

The systematic errors on branching fractions
share common contributions from the number of
produced $\psi(2S)$ events (3\%), 
uncertainty in $f$
(1.5\%), trigger efficiency (1\%), electron veto (0.5\% per veto), and MC statistics (2\%).
Other sources of uncertainty vary by channel; listed with their
contribution to the systematic error, they stem from cross-feed subtractions 
(the change induced by $\pm$50\% cross-feed variation),
accuracy of MC-generated polar angle and mass distributions 
(10\% for $b_1\pi$, 14\% for $\pi^+\pi^-\pi^0$),
and imperfect modeling of charged particle 
tracking (1\% per track), 
$\pi^0$, $\eta$ and $K_S^0$ finding (2\% per $\pi^0$ or $\eta$, 5\% per $K_S^0$), 
$\pi^\pm/K^\pm$ identification (3\% per identified $\pi/K$), and mass resolutions (2\%). 
Cross section systematic errors include the above contributions, substituting
an uncertainty in ${\cal L}$ (3\%) for the normalization error,
and accounting for uncertainties in the effects of initial and final state
radiation (7\%).
Except for $b_1\pi$ and $\pi^+\pi^-\pi^0$, statistical errors dominate.

The $\psi(2S)$ results in Table~II are consistent with
previous measurements~\cite{PDG}, where available. Unlike other VP channels, 
the isospin-violating modes $\omega\pi$ and $\rho\eta$ are not strongly suppressed 
with respect to the 12\% rule, an important new 
piece of the $\rho\pi$ puzzle.
The ratio ${\cal B}(\psi(2S)\to K^{*+}K^-)/{\cal B}(\psi(2S)\to
K^{*0}\bar{K^0}) = 0.14_{-0.06}^{+0.08}$ is found to be much smaller than
the equivalent ratio for $J/\psi$~decays, $1.19\pm0.15$~\cite{PDG}.
Fig.~4 shows that 
$\psi(2S)\to \pi^+\pi^-\pi^0$ decays have not only a distinct $\rho\pi$ 
component above the continuum contribution, but,
unlike $J/\psi\to\pi^+\pi^-\pi^0$~\cite{BESRHOPI}, which is dominated 
by $\rho\pi$, also feature a much larger cluster of events 
near the center of the Dalitz plot. 
The $\rho\pi$ results reported here do not account for
any cross-feed from this non-$\rho\pi$ component due to
its uncertain source and shape.
If five events inside the
$\rho$ mass window were attributed to the higher mass structure,
the $\rho\pi$ branching
fraction would decrease by a quarter and its significance by one unit.

The SU(3) expectation~\cite{RATIOS} for continuum cross sections is
$\omega\pi : \rho\eta : K^{*0}\bar{K^0} : \rho\pi : \phi\eta : 
K^{*+}K^- : \omega\eta : \phi\pi = 1:{2/3}:{4/9}:{1/3}:
{4/27}:{1/9}:{2/27}:0$,
in which a mixing angle $\theta_p$ satisfying
$\sin\theta_p=-{1/3}$, $\cos\theta_p={{2\sqrt{2}}/3}$ is chosen to 
describe $\eta$--$\eta'$ mixing.
With the striking exception of $K^{*0}\bar{K^0}$, 
the measured VP continuum cross sections are consistent with Born-level
calculations~\cite{WMYINTER,WMYFORM} and the above ratio predictions
from SU(3). A least-squares fit for the common unit of cross
section (corresponding to $\sigma(\omega\pi)$), excluding
$K^{*0}\bar{K^0}$, yields $\sigma_{\rm fit}=16.4 \pm 2.7$~pb with 
$\chi^2 = 4.9$ for 6 d.o.f.; $\sigma(K^{*0}\bar{K^0})$ exceeds 
$(4/9)\sigma_{\rm fit}$ by 3.0 standard deviations. 
Variations in $\theta_p$ of $\pm10^\circ$ 
induce changes of $^{+0.7}_{-1.2}$~pb in $\sigma_{\rm fit}$.

  In summary, we have presented first evidence for $\psi(2S)$
decays to $\pi^+\pi^-\pi^0$, $\rho\pi$, $\rho\eta$, and   % $\omega\pi$, 
$K^{*0}\bar{K^0}$. Measurements for several other VP channels are also given.
The results suggest that, for VP final states, $\psi(2S)$ decays
through three gluons are severely suppressed with respect to the 12\%
rule and the corresponding electromagnetic processes are not.
The decay $\psi(2S)\to \pi^+\pi^-\pi^0$ exhibits a $\rho\to\pi\pi$ signal
but has a much larger component at higher $\pi\pi$ mass.
Continuum $e^+e^-$ cross sections for these final states are 
presented for the first time.

We gratefully acknowledge the effort of the CESR staff 
in providing us with
excellent luminosity and running conditions.
This work was supported by 
the National Science Foundation and
the U.S. Department of Energy.

\

\noindent {\it Note added in proof}. Subsequent to the submission of
this Letter, similar results from BES~\cite{BESVP}
became available. After correction for
relative efficiencies and normalizations, the yields of events from $\psi(2S)$ and
continuum datasets in the BES analyses
are statistically consistent (within $\pm 1\sigma$) 
with those presented here.

\begin{table}[htp]
\setlength{\tabcolsep}{0.51pc}
\label{tab:tab1}
\caption{
For each mode: the efficiency, $\epsilon$; 
for $\sqrt{s}$=3.67~GeV data, the number of events, $N_c$,
and background from sidebands, $N_{cb}$;
for $\psi$(2S) data, the number of events, $N_{2S}$,
the estimated continuum background, $fN_c$, and
background from other $\psi(2S)$ decays, $N_b$; and
the statistical significance $S$ of the $\psi(2S)$ signal in units of a Gaussian
standard deviation. }  
\begin{center}
\begin{tabular}{lrrrrrrc}\hline\hline
Mode         & $\epsilon$(\%)& $N_c$ & $N_{cb}$ & $N_{2S}$ & $fN_c$ & $N_b$  & $S$ \\\hline
$\pi^+\pi^-\pi^0$       & 33.5 & 85 & 14 & 219 & 23.0 & 2.3&$>$6 \\ \hline
$\rho\pi$               & 28.8 & 47 &  7 &  36 & 12.8 & 1.6& 4.0 \\
\ \ $\rho^0\pi^0$       & 31.0 & 21 &  4 &  15 &  5.6 & 0.6& 2.7 \\
\ \ $\rho^+\pi^-$       & 27.7 & 26 &  3 &  21 &  7.1 & 1.0& 3.3 \\
$\omega\pi$             & 19.1 & 55 &  9 &  31 & 14.7 & 1.9& 2.9 \\
$\phi\pi$               & 15.8 &  3 &  2 &   1 &  0.8 & 1.5& $<$1\\
$\rho\eta$              & 19.5 & 38 &  2 &  29 & 10.2 & 0.9& 3.9 \\
$\omega\eta$            & 10.2 &  3 &  0 &   1 &  0.8 & 1.5& $<$1\\
$\phi\eta$              &  9.4 &  3 &  0 &   9 &  0.8 & 2.4& 2.1 \\
$K^{*0}\bar{K}^0$       &  8.7 & 36 &  2 &  35 &  9.7 & 0.5& 5.1 \\
$K^{*+}K^-$             & 16.7 &  4 &  2 &  11 &  1.1 & 3.3& 2.2 \\\hline
$b_1\pi$                & 10.9 & 22 &  4 & 288 &  5.8 &70.0&$>$6 \\
\ \ $b_1^0\pi^0$        &  6.2 &  5 &  2 &  55 &  1.3 & 9.0&$>$6 \\
\ \ $b_1^+\pi^-$        & 13.2 & 17 &  2 & 233 &  4.5 &58.0&$>$6 \\
\hline\hline
\end{tabular}
\end{center}
\end{table}

\begin{table}[htp]
\setlength{\tabcolsep}{0.37pc}
\label{tab:tab2}
\caption{
For each final state $X$: the branching fraction
${\cal B}(\psi(2S)\to X)$, with
statistical (68\% CL) and systematic errors;
the upper limit (90\%~CL), UL, for ${\cal B}$; $Q$ from Eq.~(1);
and $\sigma$, the $e^+e^-\to X$ Born-level cross section at $\sqrt{s}$=3.67~GeV. }  
\begin{center}
\begin{tabular}{lcrcr}\hline\hline
     &                 & UL        &    $Q$      & \\
Mode & ${\cal B}$~($10^{-6}$) & $(10^{-6})$ & $(10^{-2})$ & $\sigma$~(pb)\ \ \ \ \ \\ \hline
$\pi^+\pi^-\pi^0$      & $188_{-15}^{+16}\pm 28$      & 239& 7.0$\pm$1.3 & $12.3^{+1.9}_{-1.7}\pm 2.7$ \\\hline
$\rho\pi$              & $24_{-7}^{+8}\pm 2$          & 38 & 1.5$\pm$0.5 & $8.0^{+1.7}_{-1.4}\pm 1.1$ \\
\ \ $\rho^0\pi^0$      & $9_{-4}^{+5}\pm 1$           & 17 & 1.7$\pm$1.1 & $3.1^{+1.1}_{-0.9}\pm 0.5$ \\
\ \ $\rho^+\pi^-$      & $15_{-6}^{+7}\pm 2$          & 27 & 1.4$\pm$0.7 & $4.9^{+1.4}_{-1.1}\pm 0.7$ \\
$\omega\pi$            & $25_{-10}^{+12}\pm 2$        & 44 & 46$\pm$24   & $14.0^{+2.7}_{-2.3}\pm 2.0$ \\
$\phi\pi$              & $-$\ \ \ \ \                 &  7 &     $-$     & $0.2^{+1.3}_{-0.2}\pm 0.1$ \\
$\rho\eta$             & $30_{-9}^{+11}\pm 2$         & 48 & 122$\pm$49  & $10.6^{+2.2}_{-1.9}\pm 1.7$ \\
$\omega\eta$           & $-$\ \ \ \ \                 & 11 & $<$6.1      & $1.7^{+1.7}_{-0.9}\pm 0.1$ \\
$\phi\eta$             & $20_{-11}^{+15}\pm 4$        & 49 & 24$\pm$19   & $1.9^{+1.9}_{-1.0}\pm 0.2$ \\
$K^{*0}\bar{K}^0$      & $92_{-22}^{+27}\pm 9$        &141 & 17$\pm$6    & $22.4^{+4.7}_{-4.0}\pm 2.7$ \\
$K^{*+}K^-$            & $13_{-7}^{+10}\pm 3$         & 31 & 2.0$\pm$1.6 & $0.7^{+1.3}_{-0.6}\pm 0.7$  \\\hline
$b_1\pi$               & $642_{-56}^{+58}\pm 135$     &874 & 95$\pm$26   & $9.4^{+3.2}_{-2.6}\pm 2.2$ \\
\ \ $b_1^0\pi^0$       & $235_{-42}^{+47}\pm 40$      &346 & 80$\pm$30   & $2.7^{+3.4}_{-1.9}\pm 2.0$ \\
\ \ $b_1^+\pi^-$       & $418_{-42}^{+43}\pm 92$      &579 &109$\pm$33   & $6.5^{+2.4}_{-1.8}\pm 1.2$ \\
\hline\hline
\end{tabular}
\end{center}
\end{table}

\begin{figure}[thp]
\includegraphics*[width=6.5in]{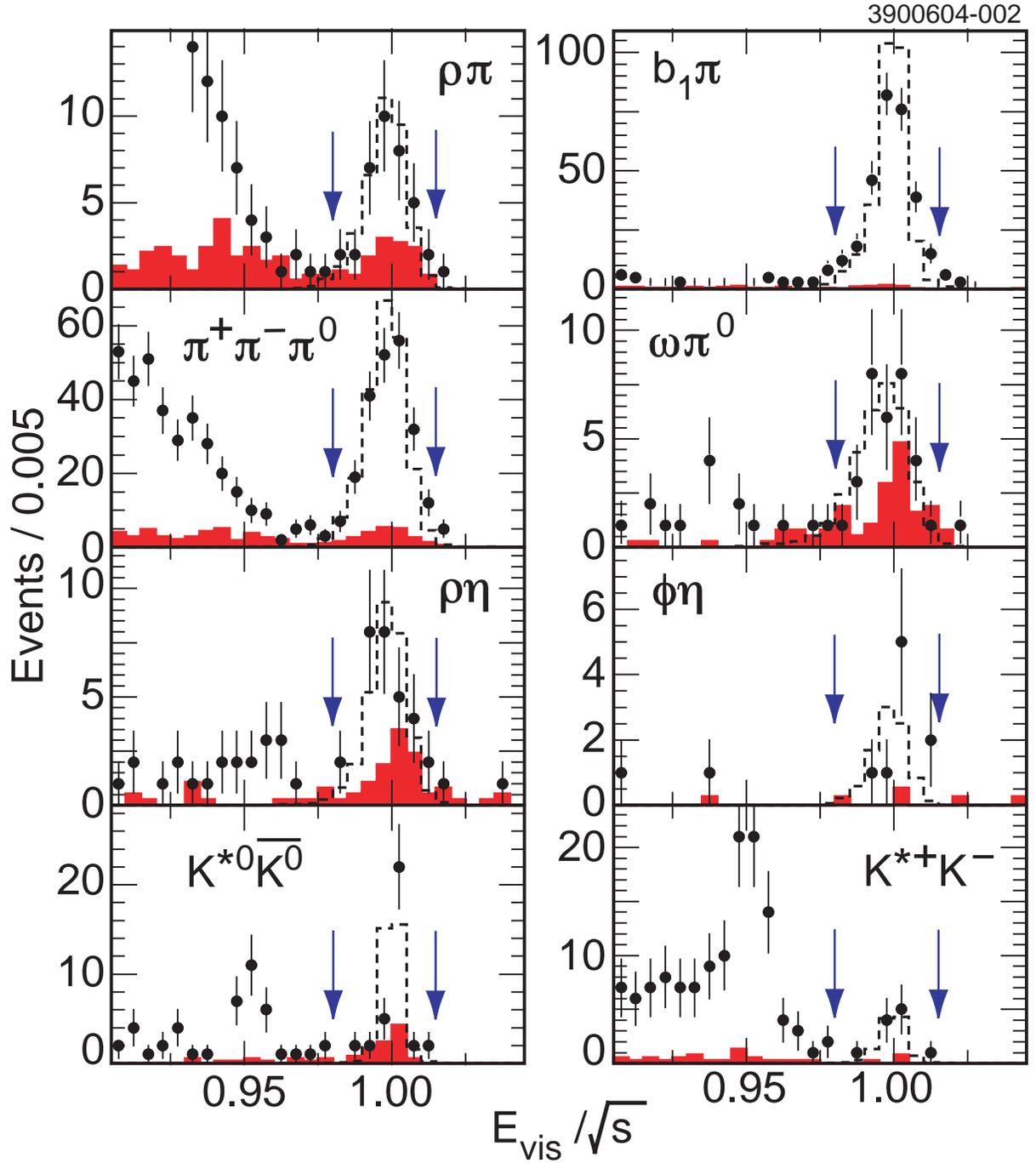} 
\caption{Distributions of scaled visible energy, $E_{\rm vis}/\sqrt{s}$, for
labeled final states. 
Plots for $\rho\pi$ and $b_1\pi$ sum over the charged and neutral states.
Histogram entries are shown for $\psi(2S)$ data (points with
error bars), scaled continuum (shaded histogram), and 
MC (dashed) with arbitrary normalization. 
The vertical arrows mark ends of signal selection ranges.}
\label{fig:fig1}
\end{figure}

\begin{figure}[thp]
\includegraphics*[width=6.5in]{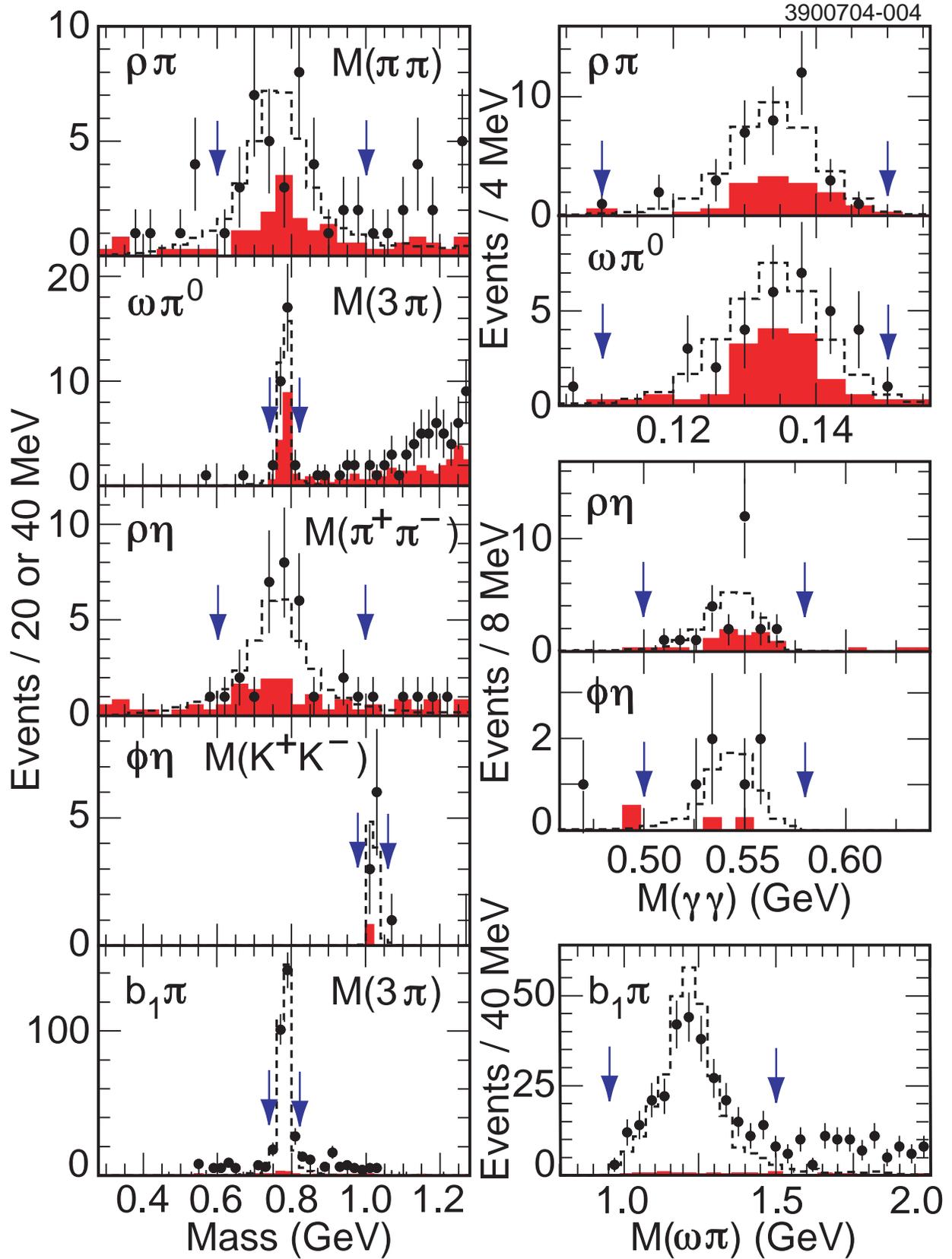} 
\caption{Invariant mass distributions relevant to the final states
indicated, one entry per event.
Symbols are defined in Fig.~1.
}
\label{fig:fig2}
\end{figure}

\begin{figure}[th]
\includegraphics*[width=6.5in]{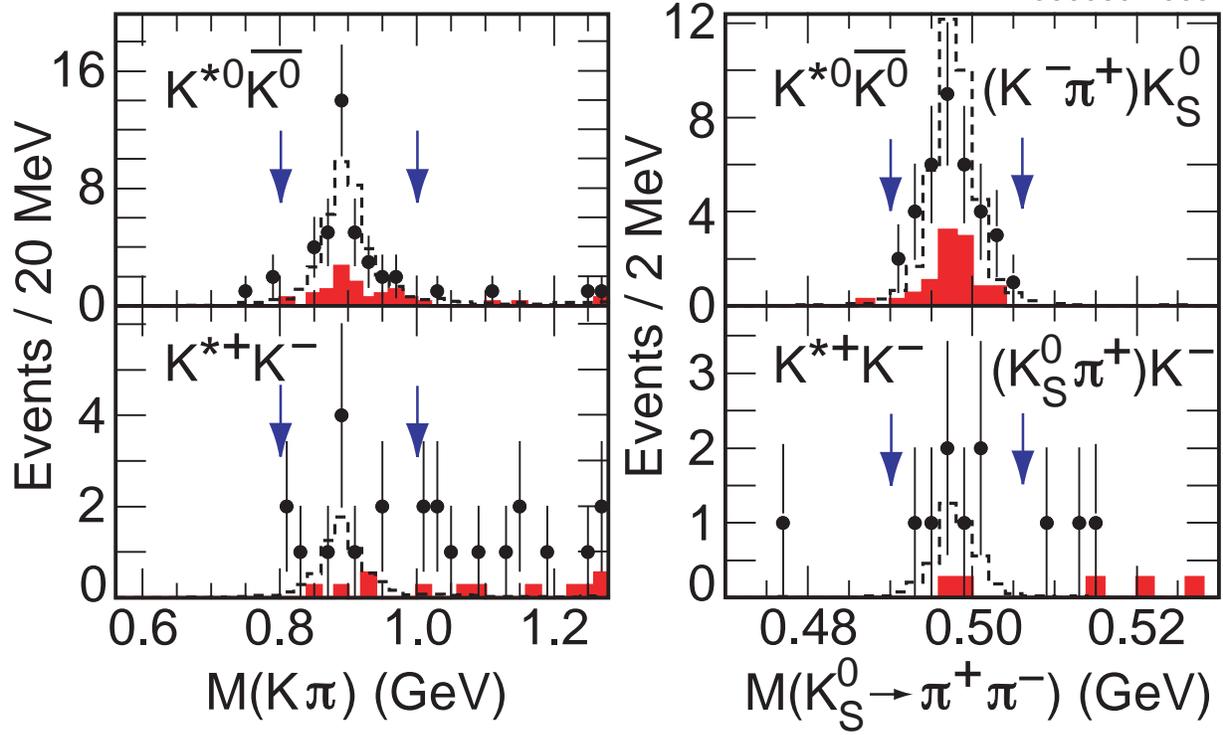} 
\caption{Invariant masses 
for the $K^*(892)$ (left) and $K_S^0$ (right) candidates,
one entry per event, for the final states indicated. Symbols are defined in Fig.~1.
}
\label{fig:fig3}
\end{figure}

\begin{figure}[thp]
\includegraphics*[width=6.5in]{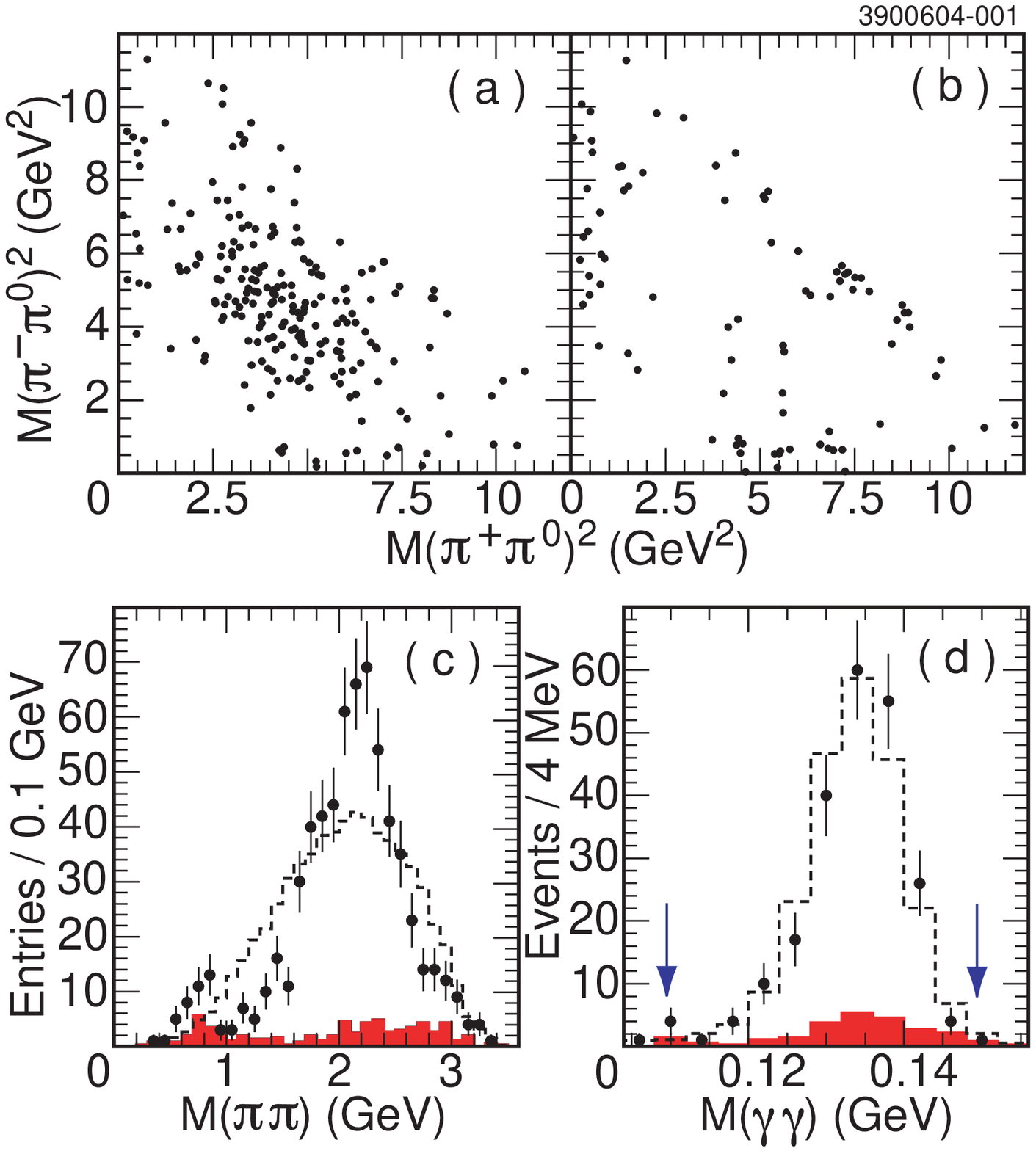} 
\caption{Distributions for the $\pi^+\pi^-\pi^0$ final state:
the Dalitz plots for $(a)$ $\psi(2S)$ and $(b)$ continuum data;
$(c)$ the $\pi^+\pi^-$, $\pi^+\pi^0$, and $\pi^-\pi^0$ mass combinations 
(3 entries/event),
and $(d)$ the $\gamma\gamma$ mass. 
Symbols are defined in Fig.~1.}
\label{fig:fig4}
\end{figure}


\begin{thebibliography}{99}

\bibitem{PDG} Particle Data Group, S.~Eidelman {\sl et al.},
Phys. Lett. B {\bf 592}, 1 (2004).

\bibitem{GULI} Y.F.~Gu and X.H.~Li, Phys. Rev. D {\bf 63}, 114019 (2001).

\bibitem{BESVT}  BES Collaboration, J.Z.~Bai {\it et al.}, Phys. Rev. D
 {\bf 69} 072001 (2004).

\bibitem{CONTIN}  P.~Wang, C.Z.~Yuan, and X.H. Mo, Phys. Rev. D {\bf 69},
 057502 (2004).

\bibitem{BRODLEP} S.J.~Brodsky and G.P.~Lepage, Phys. Rev. D {\bf 24}, 2848 (1981).

\bibitem{WMYINTER} P.~Wang, C.Z.~Yuan, and X.H. Mo, Phys. Lett. {\bf
B574}, 41 (2003).
  
\bibitem{cleoiiidetector} CLEO Collaboration,
Y. Kubota {\sl et al.}, Nucl. Instrum. Methods Phys. Res., Sect. A {\bf 320}, 66 (1992);
D. Peterson {\sl et al.}, Nucl. Instrum. Methods Phys. Res., Sect. A {\bf 478}, 142 (2002); 
M.~Artuso {\it et al.}, Nucl. Instrum. Methods Phys. Res., Sect. A {\bf
502}, 91 (2003).

\bibitem{YELLOWBOOK} CLEO-c/CESR-c Taskforces \& CLEO-c Collaboration,
 Cornell University LEPP Report No. CLNS~01/1742 (2001) (unpublished).

\bibitem{LUMINS} CLEO Collaboration, G.~Crawford {\it et al.},
 Nucl. Instrum. Methods Phys. Res., Sect. A {\bf 345},
 429 (1992).

\bibitem{BBY} C.M.~Carloni~Calame {\sl et al.}, hep-ph/0312014 [in
Proceedings of the Workshop on Hadronic Cross Section at Low Energy
(SIGHAD03), Pisa, Italy, 2003 (to be published)].
 
\bibitem{GEANT} R.~Brun {\sl et al.}, GEANT~3.21, CERN Program Library
Long Writeup W5013 (1993), unpublished.

\bibitem{BESOMEGAPIPI} BES Collaboration, M.~Ablikim {\it et al.},
Phys. Lett. {\bf B598}, 149 (2004).

\bibitem{EVTGEN} D.J. Lange, Nucl. Instrum. Methods Phys. Res., Sect. A
{\bf 462}, 152 (2001).

\bibitem{PHOTOS} E.~Barberio and Z.~Was, Comput. Phys. Commun. {\bf 79},
291 (1994).

\bibitem{ATHAR} CLEO Collaboration, S.B.~Athar {\it et al.},
Phys. Rev. D {\bf 70}, 112002 (2004).

\bibitem{BESRHOPI} BES Collaboration, J.Z.~Bai {\it et al.},
Phys. Rev. D {\bf 70}, 012005 (2004).

\bibitem{RATIOS} H.E.~Haber and J.~Perrier, Phys. Rev. D {\bf 32}, 2961
 (1985); L.~Kopke and N.~Wermes, Phys. Rep. {\bf 174}, 67 (1989).

\bibitem{WMYFORM} P.~Wang, X.H. Mo, and C.Z.~Yuan, Phys. Lett. {\bf
B557}, 192 (2003).

\bibitem{BESVP} BES Collaboration, M.~Ablikim {\it et al.},
hep-ex/0407037; hep-ex/0408047; Phys. Rev. D {\bf 70}, 112003 (2004);
 {\bf 70}, 112007 (2004);
%hep-ex/0408118; hep-ex/0410031.

\end{thebibliography}
\end{document}